\documentclass{appolb}
\usepackage{epsfig}

\begin{document}
\title{Multiplicities and Correlations at LEP%
\thanks{Invited talk presented at 
the Xth International Workshop on Deep
Inelastic Scattering (DIS2002), Cracow, 30 April -- 4 May 2002
}%
}
\author{Edward K. G. Sarkisyan
\address{CERN and University of Antwerpen}
}
\maketitle
\vspace*{-.5cm}
\begin{abstract}
A brief review on recent charge multiplicity and correlation measurements
at LEP
is given. The measurements of unbiased gluon jet multiplicity are
discussed.
Recent results on charged particle Bose-Einstein and Fermi-Dirac
correlations at LEP1 are reported.  
New results on two-particle correlations
of neutral pions are given. 
Correlations of
more than two particles (high-order correlations) obtained using
different methods are performed. 
Recent Bose-Einstein correlation measurements
at LEP2 are discussed.
\end{abstract}

\section{Introduction}

The number of hadrons, or 
{\it multiplicity}, is one of the most important observables in particle
production processes \cite{hmrep}. The distribution of
multiplicity 
is a sensitive characteristics of a
collision event. However, the multiplicity distribution tells us
just about
average, integrated numbers, while deeper information comes
from 
moments of the  distribution, which measure 
particle {\it correlations}, i.e. probe the dynamics of
the
interaction \cite{cfrep}.  

Here, I report on recent results on multiplicity and
correlation measurements at LEP. The statistics of hadronic
events
collected by each of four CERN LEP Collaborations at the Z$^0$ peak 
exceeds four million
events
and gives an unique opportunity to study  details of the theory of strong
interactions, quantum chromodynamics (QCD), and 
its applicability to (``soft'') hadron production processes.
Understanding of correlations at LEP2 is also crucial for ongoing 
Standard Model measurements.

\section{Definitions and notations \cite{hmrep,cfrep}}

The multiplicity distribution, or density $\rho_n$, of $n$ particles
with kinematic  variables $p_1,p_2,\ldots, p_n$ is defined by 
inclusive probability spectrum,

$$
\nonumber
\rho_n(p_1,p_2,\ldots, p_n)=\frac{1}{N_{\rm ev}}\,
\frac{dn(p_1,p_2,\ldots, p_n)}{dp_1 dp_2\cdots dp_n}\:, 
$$

\noindent
where $N_{\rm ev}$ is the number of events. 

As it follows from this formula, the single particle distribution
$\rho_1(p_1)$ gives an {\it average} multiplicity,
$\int
\rho_1(p_1) dp_1=\langle n\rangle\:, 
$
while integration of the $q$-particle density leads to the unnormalised
$q$th order {\it
factorial
moments},
\vspace*{-.4cm}

$$
\hspace*{-.2cm}
\int
\rho_1(p_1,p_2\ldots p_n) dp_1dp_2\cdots dp_n=\langle n(n-1)\cdots
(n-q+1)\rangle \equiv 
\langle n^{[q]}\rangle=f_q\:.
$$

\noindent 
The  normalised moments, $F_q=f_q/\langle n\rangle ^q$  have been
extensively used to  study the intermittency phenomenon \cite{cfrep}.

The $q$-particle densities give us a way to study particle 
correlations  described by $q$-particle correlation functions, 
(factorial) {\it cumulants}, 
$C_q(p_1,\ldots , p_q)$. The cumulants  vanish whenever one of their
arguments is
statistically independent, i.e. these functions measure
{\it genuine} $q$-particle
correlations. 

The cumulants are constructed from multiplicity densities, e.g. 
\vspace*{-.5cm}

\begin{eqnarray}
\label{ncors}
C_1(p_1)=\rho_1(p_1)&,
& C_2(p_1,p_2)=\rho_2(p_1,p_2)-\rho_1(p_1)\rho_1(p_2),\\
\nonumber
C_3(p_1,p_2,p_3)=\rho_3(p_1,p_2,p_3)& 
-&\sum_{(3)}\rho_1(p_1)\rho_2(p_2,p_3)+
2\rho_1(p_1)\rho_1(p_2)\rho_1(p_3)\:.
\nonumber
\end{eqnarray}
\vspace*{-.4cm}

These functions, being properly normalised, are used to study
{\it multi}-particle correlations in different kinematic variables.
\vspace*{-.2cm}

\section{Multiplicity of unbiased gluon jet}

In this Section, I consider recent results on unbiased gluon jet
multiplicity studies 
\cite{gjo,gjd}.
This analysis provides a direct check of the QCD
multiplicity predictions 
for quark jet vs. gluon jet. The approach used
allows to select  ``unbiased'' gluon jet in 3-jet events, i.e. it is
independent of jet-finding algorithm which were usually applied in earlier
studies \cite{hmrep,gjo}.   

In theory the gluon jet multiplicity, $N_g$, 
is
defined in gluon-gluon ($gg$) jet systems, while experimentally the
$N_g$  multiplicity is obtained from 3-jet
$q{\bar q}g$ final states. To this end, one uses the formula
which connects 3-jet
multiplicity with $q{\bar q}g$ and $gg$ multiplicities,
$N_{q{\bar q}}$ and $N_{gg}$, respectively, 
\vspace*{-.3cm}

\begin{equation}
N_{q{\bar q}g}=N_{q{\bar
q}}(L,k_{\perp,Lu})+\frac{1}{2}N_{gg}(k_{\perp,Lu})\:,
\label{gglu}
\end{equation}
\begin{equation}
N_{q{\bar q}g}=N_{q{\bar
q}}(L_{q{\bar q}},k_{\perp,Le})+\frac{1}{2}N_{gg}(k_{\perp,Le})\:.
\label{ggle}
\end{equation}

\noindent
Here, $L$ specifies the e$^+$e$^-$ c.m.s. energy, while $L_{q{\bar q}}$
the 
$q{\bar q}$ system energy. The two expressions are given by two approaches
to define the gluon jet energy w.r.t. $q{\bar q}$ system by Lund and
Leningrad groups.    The important fact is that $N_{gg}$ depends only on a
single scale $k_{\perp}$, i.e. it is {\it unbiased} in contrast to 
$N_{q{\bar q}}$ depending on the energy scale too. 

Fig. 1 shows OPAL results on $N_{gg}$ of charged particles as a function
of the jet energy
$Q$ \cite{gjo}. One can see that calculations using the Lund approach
better
describe the data and Monte Carlo predictions (the latters well reproduce
the data)
than that of Leningrad.

The Lund formalism was proceed to obtain the ratios, $r^{(j)}\equiv
(d^jN_g/d\varepsilon)/$ $(d^jN_q/d\varepsilon)$ of gluon and quark
multiplicities. Here $\varepsilon$ specifies the jet energy. 
The ratios were found \cite{gjo} to satisfy the QCD prediction,
$r^{(0)}<$ $r^{(1)}<$ $r^{(2)}\to$ 2.25 as $Q\to \infty$. From this it was
obtained an effective value of QCD colour factors, $C_{\rm A}/C_{\rm F}$ = 
2.23 $\pm$ 0.14 being in a good agreement with the QCD value of 2.25. A
similar
value is preliminarly reported by DELPHI
\cite{gjd}. 

\hspace*{-.5cm}
\begin{minipage}[p]{7cm}
\vspace*{.25cm}
\centering\epsfig{figure=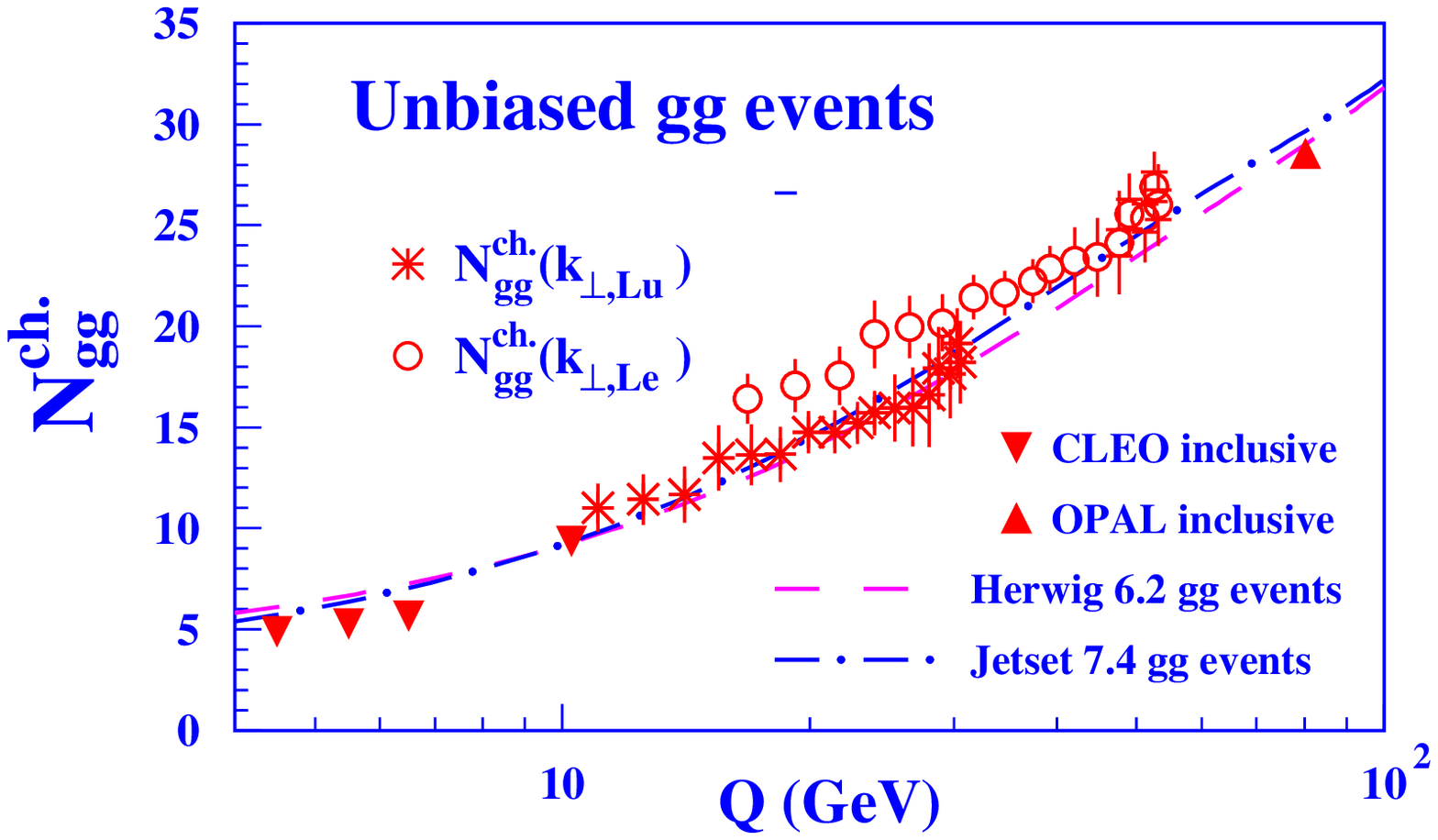,
width=8.cm}
\vspace*{.077cm}
\end{minipage}

\vspace*{-5.3cm}
\hspace*{7.7cm}
\begin{minipage}[p]{4.1cm}
\vspace*{.25cm}
{\footnotesize
{\bf Fig. 1.} The average charged particle multiplicity of unbiased $gg$
events as a function of energy scale.  Different $k_{\perp}$
correspond to Eqs. (\ref{gglu}) and (\ref{ggle}), respectively.
See text and Ref. \cite{gjo} for more
details.}  
\vspace*{.077cm}
\end{minipage}
\vspace*{.75cm}

\section{Two-particle correlations in hadronic Z$^0$ decays}

During last years, LEP Collaborations actively study two-particle
correlations of bosons, Bose-Einstein correlations (BEC), and fermions,
Fermi-Dirac correlations (FDC) \cite{wkbec}. To study
two-particle correlations one needs to measure
$\rho_2(p_1,p_2)/\rho_1(p_1)\rho_1(p_2)$, where
$p_1$ and $p_2$
are the 4-momenta of particles. Experimentally, one measures
$C(Q)=\rho_2(Q)/\rho_2^{0}(Q)$,
where $Q^2=-(p_1-p_2)^2$. The normalisation $\rho_2^{0}$ of a reference
sample has to
be free of BEC/FDC and 
can be defined in different ways: Monte Carlo without such kind of
correlations, $\rho_2(Q)$ of
unlike-sign hadrons, pairs with particles from different events, or from
different hemispheres (mixings).  Then the $C(Q)$ function fit assuming
the
Gaussian source,
\vspace*{-.4cm}

\begin{equation}
C(Q)=1\pm\lambda\,\cdot {\rm exp}(-Q^2R^2)\:, 
\label{gauss}
\end{equation}
\vspace*{-.5cm}

\noindent 
gives $\lambda$ to be a measure of the  strength of the correlations while
$R$ is
considered
as emitter radius. In this
approach $\lambda=1$ indicates completely incoherent emission. The ``+''
sign stands for BEC, and the ``$-$'' for
FDC.

BEC of {\it charged} pions are well established at LEP, and the radius is
obtained to vary between 0.5 and 1.0 fm depending on the reference sample.

Recently, L3 measured BEC of {\it neutral} pions \cite{l3pi0}. It is
found that in the
same
framework for charged and neutral pions,
$R(\pi^0\pi^0)<R(\pi^{\pm}\pi^{\pm})$ (with 2$\sigma$ evidence). This is 
 in
qualitative agreement with the Lund string model.  

A decrease of $C(Q)$ as $Q\to 0$ for fermions was observed for
$\Lambda$
pairs \cite{wkbec}, while
OPAL preliminarly reported \cite{ofdp} on a depletion in
antiproton pairs. 

Combining measurements of  emission radius of  pions, kaons, 
Lambdas and antiprotons,  the hadron
mass hierarchy
$R_\pi>R_K>R_{\Lambda,{\bar {\rm p}}}$ is obtained as shown in Fig. 2
\cite{rmass}.
This hierarchy can be
explained by
Heisenberg uncertainty principle model \cite{rmass} 
or with correlation between
space/time and momentum/energy of the particle in the hadroproduction
process \cite{rmassb}.  Meantime, the interpretation of
$R$ as the emitter radius leads to the very high emitter energy density  
at baryon mass, $\sim$100 GeV/fm$^3$ \cite{rmass}.  
\vspace*{-.1cm}

\hspace*{-1.cm}
\begin{minipage}[p]{7cm}
\vspace*{.25cm}
\centering\epsfig{figure=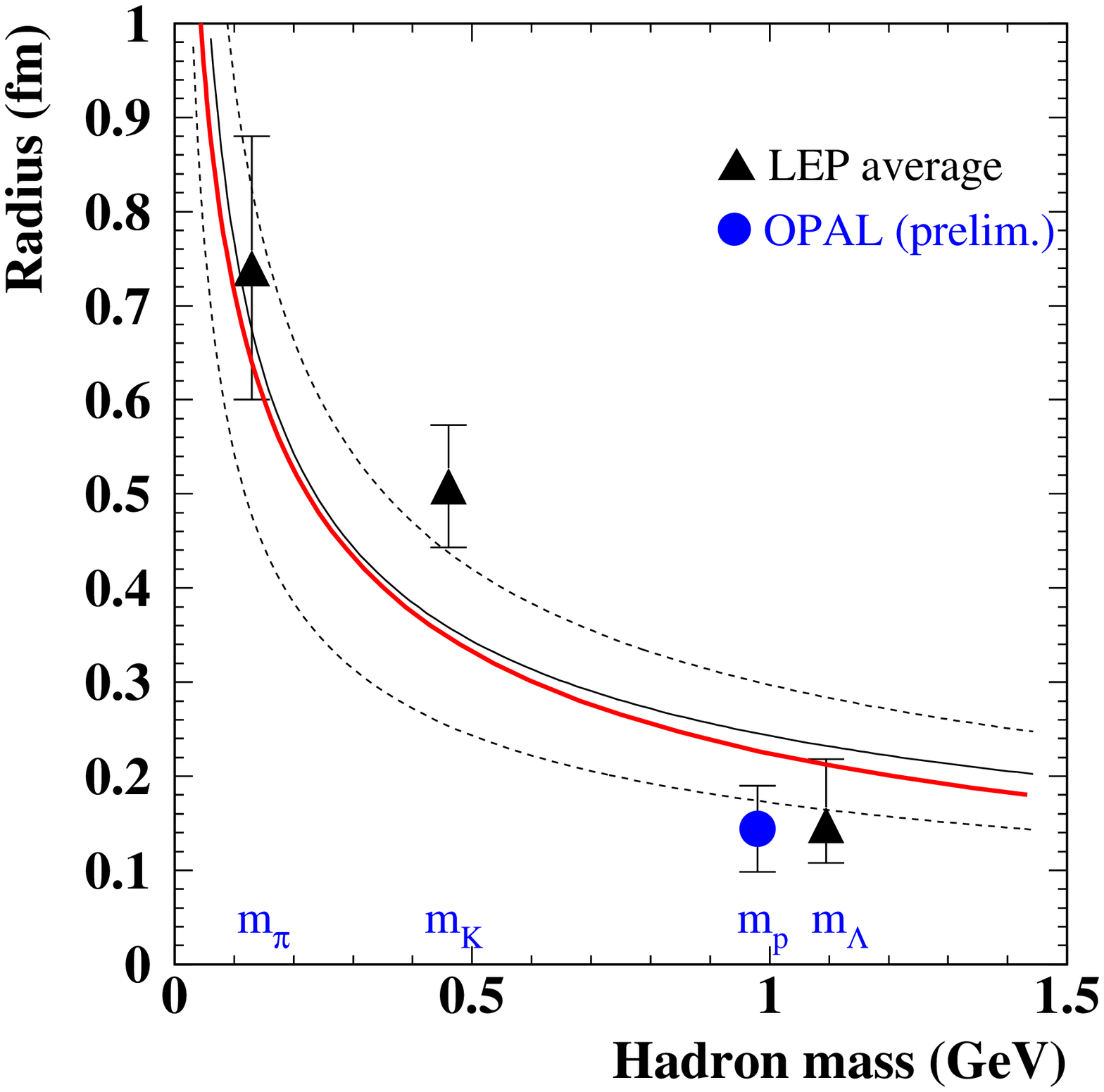,
width=5.5cm}
\vspace*{.077cm}
\end{minipage}

\vspace*{-5.3cm}
\hspace*{5.7cm}
\begin{minipage}[p]{6.2cm}
{\footnotesize {\bf Fig. 2.} The emitter radius $R$ as a function of the 
hadron mass
obtained from LEP \cite{rmass}. The thin lines are from the Heisenberg
uncertainty relations with the time scale values of 10$^{-24}$ (central
thin line) and 0.5$\cdot $10$^{-24}$ and 1.5$\cdot $10$^{-24}$ (thin
dashed
lines).
The thick line is given by the virial theorem
with a general QCD potential. For more details, see Ref. \cite{rmass}.}  
\vspace*{.077cm} 
\end{minipage}
\vspace*{.7cm}

\section{High-order correlations at the Z$^0$ peak}
\vspace*{-.1cm}

Further step in understanding the  hadroprduction process is to analyse
the correlations of more than two particles, i.e. high-order
correlations
\cite{cfrep,wkbec}. 

At LEP, 3-particle BEC study has
been 
recently performed by L3 \cite{l3bec3}, in addition to the earlier
studies \cite{wkbec}.
A key point in this analysis is to remove {\it non-genuine} 3-particle
correlations of two- and single-particle product, see $C_3$ in
Eqs.(\ref{ncors}). Then, similarly to
the two-particle case analysis is carried out using  
$Q_3$ variable along with  no-BEC normalisation $\rho_3^0(Q_3)$.     
Extended  Gaussian fit (\ref{gauss}) of the genuine 3-particle
correlation
function with Hermite polynomials 
and used the Fourier transform 
of a source density, L3 concludes with fully
incoherent pion production mechanism.

Genuine correlations up to the 4th order are observed
by OPAL for like-sign pions in terms of 
the normalised cumulants, Eqs.(\ref{ncors}) \cite{ocum}. 
BEC algorithm of PYTHIA Monte
Carlo is found to reproduce multiparticle correlations in 1- to
3-dimensional phase space regions of rapidity,
azimuthal angle and transverse momentum. Interrelation between
higher-order and two particle correlations  are obtained.  
\vspace*{-.2cm}

\section{Bose-Einstein correlations at LEP2}

In WW 4-quark hadronic decays, it is difficult to separate
products of each W decay since separation of the decay vertices is $\sim$
0.1
fm, while the hadronisation scale is of 0.5 -- 1 fm. Therefore, due to
possible {\it inter}-W BEC, the Statndard Model measurement of 
the 
W mass is  expected to be  biased and needs 
BE effect between observed hadrons to be taken into account.  

At LEP, we used the method of $\Delta \rho(Q)$ function of
two-particle densities, 
\vspace*{-.7cm}

$$
\Delta \rho=\rho_2^{\rm WW\to 4q} -(2\rho_2^{\rm W\to 2q}+\rho_2^{\rm
mix})\:, 
$$

\noindent
and $D(Q)$ being the ratio of $\rho_2^{\rm WW\to 4q}$ to the sum in
parentheses \cite{rho}.
 The latter represent faked 4q events  
of the hadronic part of semileptonic events and 
mixed events  
from two independent  
semileptonic events without leptons. If there is no inter-W BE effect,
then
 $\Delta
\rho=0$ and $D=1$. 

The $\Delta
\rho(Q)$ measurements show consistency with the absence of the inter-W
BEC, and $D$ is
found to be about 1 \cite{rlep}. Further study is ongoing.   
\medskip

\noindent
I am thankful to DIS2002 Org. Committee for inviting me, 
to my colleagues from LEP for their kind help and support. I
apologise to those whose results were not covered due to the lack of time
and space available for this talk.

\end{document}